\documentclass[nohyper,12pt,letterpaper]{JHEP}
\usepackage{epsfig}

\newfont{\frak}{eufm10 scaled 1200}

\newfont{\Bbb}{msbm10 scaled 1200}     
\newcommand{\mathbb}[1]{\mbox{\Bbb #1}}
\DeclareSymbolFont{AMSa}{U}{msa}{m}{n}
\DeclareSymbolFont{AMSb}{U}{msb}{m}{n}
\let\Box\relax
\DeclareMathSymbol{\Box}{\mathord}{AMSa}{"03}

\def \eqn#1#2{\begin{equation}#2\label{#1}\end{equation}}

\title{A Model for High Energy Scattering in Quantum Gravity }

\author{Tom Banks\\
  Department of Physics and Astronomy\\
  Rutgers University, Piscataway, NJ 08855-0849\\
E-mail: \email{banks@physics.rutgers.edu}}
\author{Willy Fischler\\
  Department of Physics\\
  University of Texas, Austin, TX 78712\\
E-mail: \email{fischler@mail1.ph.utexas.edu}}

\abstract{
We present a model for high energy two body scattering in the
quantum theory of gravity.  The model is applicable for center of mass
energies higher than the relevant Planck scale.  At impact parameters
smaller than the Schwarzchild radius appropriate to the center of mass
energy and total charge of the initial state, the cross section is
dominated by an inelastic process in which a single large black hole is
formed. The black hole then decays by Hawking radiation.  The elastic
cross section is highly suppressed at these impact parameters because of
the small phase space for thermal decay into a high energy two body state.
For very large impact parameter the amplitude is dominated by eikonalized
single graviton exchange.  At intermediate impact parameters the
scattering is more complicated, but since the Schwarzchild radius grows
with energy, we speculate that a more sophisticated eikonal calculation
which uses the nonlinear classical solutions of the field equations
may provide a good approximation at all larger impact parameters.
We discuss the extent to which black hole production will be
observable in theories with low scale quantum gravity and large dimensions.

}


\preprint{\hepth{9906038}\\RU-99-23, UTTG-03-99}
\begin{document}


\section{Introduction}
The problem of very high energy scattering is deeply intertwined with
the history of string theory and the quantum theory of gravity.  Indeed,
string theory was originally invented as a model of the Regge behavior
expected for high energy scattering in hadron physics.  More recently,
there has been a great deal of effort devoted to elucidating the
behavior of high energy scattering in string theory
\cite{hienstringa}-\cite{hienstring} .  In parallel to this activity,
't Hooft and followers \cite{thverl}, have initiated a study of high
energy scattering in the `` quantum theory of gravity '', which is to
say that their considerations are supposed to be valid in any theory
which obeys the principles of quantum theory and reduces to General
Relativity at long distances.  The aim of this latter program was, at
least in part, to address the question of information loss in black hole
formation.  Not coincidentally, the work cited in
\cite{hienstringa}-\cite{hienstring} restricts attention to impact
parameters larger than those at which General Relativity would predict
that black holes are formed.

The claim of the present note is that
the gross features of high energy scattering far above the Planck
scale can be extracted from semiclassical considerations in General
Relativity.  We will always consider situations with four or more
asymptotically Minkowski dimensions.
For simplicity, we will restrict our attention to initial
scattering states consisting of two particles of mass far below the
Planck scale, but it should be possible to extend it to more complicated
initial states.
The basis of this claim is the following simple
observation.  The classical picture of this initial state, insofar as
gravitational interactions are concerned, consists of two
Aichelburg-Sexl shock wave metrics.  General Relativity predicts that
when the impact parameter of the two shock waves is smaller than a
critical value $R_S$, a black hole is formed.  The No-Hair theorem tells
us that the classical final state is then uniquely specified by its
representation under the asymptotic symmetry group. $R_S$ is of order
the Schwarzchild radius of the corresponding black hole and we will, by
abuse of language, call it the Schwarzchild radius.  The mass of the
black hole is of order the center of mass energy of the collision.
Thus, $R_S$ grows with the center of mass energy.

On the other hand, for asymptotically large impact parameters,
scattering is also described by classical General Relativity.
Indeed, all existant
calculations are consistent with the claim that high energy large impact
parameter scattering is dominated by eikonalized single graviton exchange.
Thus, since $R_S$ grows with energy, we may expect that all aspects of
the scattering up to the point of formation of the black hole are well
described by the classical theory.  Unfortunately, except in the case of
$2+1$ dimensions with Anti-deSitter boundary conditions \cite{ads}
the exact classical solution for black hole formation in the collision
of Aichelburg-Sexl shock waves is unknown.  The state of the art
calculations for shock wave initiated processes in four dimensional flat
space may be found in \cite{death}.  This fact will make it impossible for us
to exhibit a complete formula for scattering cross sections at all
energies and impact parameters.

For impact parameters below $R_S$ an exact description of scattering
amplitudes would require us to enter into all of the mysteries of the
black hole information problem.  However, since for large energy the
mass of the black hole is large, one needs only the familiar Hawking
formulae to extract the gross features of inclusive cross sections.
Furthermore, the thermal nature of these cross sections suggests that
any more precise description of the scattering will be hopelessly
complicated.
We want to emphasize that, although we believe recent progress in
M~Theory suggests that the scattering matrix is unitary even in the
presence of black hole production, our results do not depend heavily on
this assumption since we only describe inclusive cross sections.

To summarize then, our proposed model for high energy scattering is the
following: At impact parameters greater than $R_S$ elastic and
inelastic processes (gravitational radiation, photon bremstrahlung for
charged particles and the like) will in principle be described by
solving the classical equations of the low energy theory with initial
conditions described by a pair of shock waves with appropriate quantum
numbers\footnote{Note that we only attempt to describe the leading high
energy behavior.  The full series of corrections to this behavior would
require more knowledge of M~Theory than we possess.} .  Note that since
$R_S^2$ grows much faster with energy than strong interaction cross
sections are allowed to (by the Froissart bound), this behavior will be
completely determined by the classical physics of the degrees of freedom
with energies below the Planck scale (see however the discussion of
weakly coupled string theory below).

At smaller impact parameters, scattering will be dominated by
``resonant'' production of a single black hole with mass equal to the
center of mass energy.  We put the word ``resonant'' in quotes because,
despite their long lifetimes, black holes do not fit the profile of a
classic Breit-Wigner resonance.  Indeed they are most peculiar from the
Breit-Wigner point of view.  The Breit-Wigner formula for the
contribution of a particular resonance to the elastic cross section for
two body scattering is proportional to the square of the partial width
for the resonance to decay into this particular channel.  This is
because, for a single narrow resonance, unitarity implies that the
amplitude to produce the resonant state is the same as that for the
resonance to decay back into the initial state.  For black hole
production, we expect the initial amplitude to be of order one whenever
the impact parameter is smaller than $R_S$.  On the other hand, the
decay of the black hole is thermal; there will be a very small
probability for it to decay back into the initial high energy two
particle state.  Thus, we expect the elastic cross section to be linear
rather than quadratic in the partial width of the two body final state.
Thus, the elastic cross section is larger than might have been expected.

More striking is the inelastic resonant cross section, $\sigma (A + B
\rightarrow BH \rightarrow Anything)$.  This will be large even though
the partial width to decay into the initial state is small.  The cross
section resembles that for a high energy collision of two bodies with an
already existing, highly degenerate, macroscopic object.  In such a
situation, the energy of the initial particles is thermalized among the
large number of degrees of freedom of the macroscopic body, the decay is
thermal, and the probability of recreating the initial state much
smaller than that of the initial collision.

A. Rajaraman has suggested to us that a fact from classical GR may help
to explain this behavior. When a black hole is formed by the collapse of
a thin spherical shell of matter, the horizon forms long before the
shell has fallen past the Schwarzchild radius.  Similarly, in the
collision of two shock waves, we might expect the horizon to form long
before the waves reach the distance $R_S$.  In this sense we can view
the scattering as being caused by the impact of the colliding particles
with an already existing horizon, an object which has a macroscopic
number of degrees of freedom.

The other important difference between black hole and resonance
production is that resonances occur at discrete energies.  By contrast,
for every center of mass energy $E$ above some threshold of order the
Planck mass, and every impact parameter below $R_S (E)$ we expect the
high energy cross section to be dominated by the production of a single
object, almost stationary in the center of mass frame.  The object will have a
long lifetime and will decay thermally (and thus isotropically in the
center of mass frame) .  The elastic cross section will be small.
Note in addition that two body final states
with large momentum transfer will be even more highly suppressed than
the generic two body state.  This is in
marked contrast with the behavior of the system below the black hole
production threshold, where the proliferation of hadronic jets in the
final state increases with energy.  Once black holes 
with large enough radii can be formed, the
colliding particles never get close enough to perform a hard QCD
scattering.

The dramatic nature of these processes suggests that they will be easy
to see if we ever build a Planck energy accelerator.  This is
particularly exciting in view of recent suggestions that the world may
have large extra dimensions and a true Planck scale of order a TeV
\cite{largedim}\footnote{We note that the authors of \cite{savarg} have
investigated the properties of and astrophysical constraints on, black
holes in theories with large extra dimensions.} . 
However, we argue that most of the Hawking radiation
of the higher dimensional black hole will, for phase space reasons,
consist of Kaluza-Klein modes of gravitons, and thus be undetectable.
In the absence of experimental information about the final state, it
is hard to distinguish these missing energy signals from those which
come from production of a few KK gravitons \cite{kkdetect}.  At
sufficiently high energies the Hawking temperature of the black hole is
small compared to the KK energy scale and the Hawking radiation will be
dominated by observable particles.  We show this occurs at about the
point where the Schwarzchild radius is equal to the KK radius, which is
at energies above the four dimensional Planck mass.

We suggest that the suppression of hard QCD processes is a possible
signal for identifying this sort of invisible black hole production.
Complete suppression of QCD jet phenomena requires that the Schwarzchild
radius be larger than an inverse GeV, and this only occurs at
unreachably high energies.  However, the suppression of jets with
transverse momenta higher than the inverse Schwarzchild radius should
become apparent before this.   The detailed investigation of this
phenomenon is beyond the scope of the present paper. 

The plan of the rest of this paper is as follows.  In the following
section we outline the regime of parameters within M~Theory in which we
might expect our discussion to be valid.  We point out in particular
that physics in the regimes corresponding to weakly coupled string
theories is considerably more complex than what we have discussed.
There can be a plethora of scales and a variety of different high energy
regimes.  Readers of a more phenomenological bent are advised to skip
this section, which will be of interest mostly to string theorists.
In Section 3, we present some remarks relevant to the regime
discussed in this introduction, and assess the likelihood of observing
black hole production experimentally if theories with large dimensions
are correct.

\section{High Energy Scattering in Weakly Coupled String Regimes}

Consider the moduli space of M~Theory compactifications with
four Minkowski dimensions.  Much of this moduli space can be well
approximated by compactifications of 11 dimensional supergravity (11D
SUGRA) on manifolds with dimensions large compared to the Planck scale.
The discussion in the introduction applies primarily to such
regions of moduli space.  An example of regions not covered by this
description are weakly coupled string compactifications.  These can be
viewed as the proper limits of compactifications of 11D SUGRA on
manifolds with some dimensions much smaller than the Planck scale.
The same is true for F theory compactifications.  We will begin by
discussing the simple case in which all dimensions are of order or much
larger than the eleven dimensional Planck length.  Then by examining
the case of weakly coupled string theory, we will establish that other
regimes have a much more complicated set of high energy behaviors.

Imagine first that some dimensions are of order the Planck scale, while
$n$ others are much larger.  Let $M_{4+n}$ denote the Planck mass in the
effective theory below the eleven dimensional Planck scale.  Given our
assumptions, it is of order the eleven dimensional Planck scale and we
will not bother to distinguish between them.  The four dimensional
Planck scale is given by
\eqn{mp}{M_P^2 \sim V_n M_{4+n}^{2+n},}
where $V_n$ is the volume of the large dimensions.   Note that $M_P >
M_{4+n}$.

As the energy is raised, we will approach two thresholds, the first, the
Kaluza Klein (KK) scale of the large dimensions, and the second,
$M_{4+n}$.  The observation of \cite{largedim} is that if $V_n$ is large
in eleven dimensional Planck units, if the standard model lives on a
brane embedded in the large dimensions, and if gravitons and other fields
with only nonrenormalizable couplings to the fields on the brane are the
only bulk fields, then the first threshold may show up only in very high
precision experiments.  The couplings of ordinary matter to the new
states will be suppressed by powers of the energy divided by $M_P$ until
we reach the threshold $M_{4+n}$.  Most analyses of what happens above
this threshold have concentrated on the production of KK modes.  We will
argue that somewhere around this energy regime, the (in principle)
much more dramatic
phenomenon of black hole production sets in.  We will reserve
a more detailed description of these processes for the next section.  Here we
merely observe that the appropriate form of GR to use in these estimates
is $4+n$ dimensional gravity.

We turn now to regimes described by weakly coupled string theory.  Here
there is generically a hierarchy of scales, starting with the string
scale and proceeding to energy scales which are larger than the string
scale by inverse powers of the coupling.  Among these is the Planck
scale associated with the $4+n$ dimensional space \footnote{We now assume that
$6-n$ dimensions are compactified at about the string scale.  In string
theory, T dualities and Mirror symmetries usually make this a lower
bound on compactification dimensions in the weak coupling regime.}.

Since we are interested in scattering above the string energy the
effective theory in the regime of interest will always be ten
dimensional (see the previous footnote).   There have been many attempts
to study high energy scattering in string perturbation theory.  We will
argue that these are reliable only in a certain energy
regime\footnote{The analysis of the next few paragraphs summarizes
unpublished work done several years ago by
 one of the authors (TB) and S. Shenker. We thank S. Shenker for
permission to include it here.}.

Let us turn first to the fixed angle regime studied in \cite{hienstringa}.  A
cartoon version of the analysis of this papers follows: All Lorentz
invariants of the scattering process have the same order of magnitude,
call it $s$, in this regime.  $k$ loop amplitudes have the form
\eqn{kamp}{A_k \sim \int dm e^{- s\alpha^{\prime} f_k (m)/k}}
where the integral is over the moduli space of genus $k$ Riemann
surfaces with $4$ punctures and $f_k$ depends only weakly on $k$.
For large $s$ one does the integral by steepest descents, obtaining an
amplitude which falls off like $e^{- s \alpha^{\prime} f_k (m_0 )/k}$.
The Gaussian
fluctuations around the stationary point in moduli space give a coefficient
of order $6k\!$.

The facts that the exponential becomes flatter for large $k$ and that
the coefficient far exceeds the $2k\!$ growth expected for the large
order behavior of string perturbation theory \cite{shenker} (facts which
are mathematically related) lead us to be suspicious of this result at
energies which scale like inverse powers of $g_S$ \footnote{ Physically, the
reason for this flattening was explained in
\cite{hienstringa}: the lowest order amplitude gives an exponentially falling
amplitude for large momentum transfer.  At higher orders, the most
efficient way to distribute the momentum transfer is to form a $k$
string intermediate state, with each subprocess transferring momentum
squared of order $s/k$.  }.
Indeed, the large $k$ behavior of the amplitude at fixed $s$ is constant
in $s$ and is of order $2k\!$ (the estimate of the coefficient comes
from the volume of moduli space).  This suggests an $s$ independent,
nonperturbative
contribution to the amplitude of magnitude $e^{- {c\over g_S}}$ such as
that predicted by D-instantons in Type IIB string theory.  One may
expect similar pointlike contributions in other weakly coupled string
theories (with the notable exception of the heterotic theory where these
contributions may have something to do with the throats of NS 5 branes)
from components of the wave functions of scattering states
which contain D-object anti D-object pairs separated by distances of
order the string length.  At small impact parameter we should see a
contribution from the pointlike scattering of individual D-branes
\cite{dkps}.

The nonperturbative amplitude competes with the perturbative one when
$s\alpha^{\prime}
 \sim {1\over g_S}$.  Note that the ten dimensional Planck energy squared
is $M_{10}^2 \sim (g_S^{1/2} \alpha^{\prime} )^{-1}$, which is much smaller
than this crossover energy.  The semiclassical analysis of the
introduction and the following section are valid only when the energy is
much larger than $M_P$ and the Schwarzchild radius larger than the
impact parameter as well as the string length.
In ten dimensions the Schwarzchild radius is of order
$s^{1\over 14} g_S^{2\over 7}$ in string units, so the semiclassical
regime would apparently only set in for $s \sim g_S^{-4} $.
This would seem to be a valid estimate in IIB string theory, but in IIA
this energy is above the inverse
compactification radius of the M~Theory circle, so we
should really make an eleven dimensional estimate.  The eleven
dimensional Schwarzchild radius only exceeds the string scale when $s
\sim g_S^{-6}$ in string units.  Thus in all cases it seems that the
crossover between perturbative string and D-instanton behavior sets in
in a regime in which gravitational corrections are negligible.

If the interpretation of the pointlike nonperturbative cross section in
terms of D-brane ``sea partons'' in the string wave function is correct,
we may expect that the description we have given of the crossover is not
complete.  Indeed, the authors of \cite{dkps} showed that D0 brane
scattering in weakly coupled type IIA string theory became soft at
scales of order the eleven dimensional Planck mass, or $g_S^{-1/3}$ in
string units.  This is lower than the crossover scale.

To conclude, in the weakly coupled string regime, the semiclassical
analysis of the introduction is expected to be valid only at energies
which are parametrically (in $g_S$) higher than any relevant Planck
scale.  In the IIA theory it is only 11 dimensional SUGRA which
eventually becomes relevant, and only at an energy scale parametrically
larger than the inverse compactification radius of the M~Theory circle.
At energies below this true asymptopia we expect to
see a rich structure of high
energy amplitudes, dominated successively by perturbative strings, and
nonrelativistic followed by relativistic scattering of `` Dirichlet sea
'' constituents of the incoming states.

\section{Black hole cross sections}

We write the elastic amplitude for $2 \rightarrow 2$ scattering in
eikonal form
\eqn{eik}{A(s,q^2) \propto \int d{\bf b} e^{i \chi ({\bf b}, s)}}
where ${\bf b}$ is the impact parameter and $s$ the square of the center
of mass energy.
For $n$ relatively large compact dimensions, the Schwarzchild radius
of a $4+n$ dimensional black hole of mass $\sqrt{s}$
is approximately , $R_S \sim M_{4+n}^{-1} (s/M_{4+n}^2)^{1\over 2(n+1)}
$.  In order to use flat space black hole formulae, we must have
$R_S \ll L$ , the radius of the compact dimensions.  In terms of
the energy, this bound is $\sqrt{s} \ll M_P
(M_P / M_{4+n})^{n+2 \over n}$.  For applications to theories with
low scale quantum gravity, this bound is never exceeded, so we will
not discuss larger values of $s$.  We note however that when $R_S$
exceeds the compactification radius, the most likely outcome is that
the system is described as a four dimensional black hole (a black
brane wrapped on the compact dimensions).

For impact parameters smaller than $R_S $ the cross section will
be completely dominated by black hole production.  As outlined
in the introduction, this will have the following consequences:
\begin{itemize}

\item The elastic cross section will be suppressed by a Boltzmann
factor $e^{- \sqrt{s}/ T_H}$, where the Hawking temperature
$T_H \sim M_{4+n} (M_{4+n} / \sqrt{s})^{1\over n+1}$.

\item Due to initial state bremstrahlung the black hole will not
be exactly at rest in the center of mass frame.  The average energy
emitted in bremstrahlung should be calculable by the methods of
\cite{death}.  The final state will be a black hole at rest in the
frame determined by this bremstrahlung calculation.  It will decay
thermally, and therefore isotropically in this frame.  This prescription
only allows one to calculate inclusive cross sections, but the
thermal
nature of these indicates that any more precise calculation of the
amplitudes for various final states is beyond the range of our
abilities.

\item In the standard model, we expect high energy collisions to
be characterized by a larger and larger multiplicity of QCD jets
with higher and higher transverse momenta.  One of the most striking
features of black hole production is that processes with transverse
momenta larger than $R_S^{-1}$ should be completely absent.
The incoming particles never get close enough together to perform
a hard QCD scattering.  This characteristic shutoff of hadronic jets
may be one of the most striking signals of black hole production
processes.

\item Although a long lived black hole will be produced at every
sufficiently high energy and small impact parameter, the signature
of these events does not look like a conventional Breit-Wigner
resonance.  

\end{itemize}
When the impact parameter is larger than $R_S$, we do not expect
black hole formation to occur.  When the impact parameter is very
large,
the elastic scattering is given by the eikonal formula coming from
single graviton exchange.  Note that here it is
four dimensional gravitational physics which is relevant
since we are talking about asymptotically large impact parameter.
At energies relevant for discussing theories of low scale quantum
gravity , these amplitudes are completely negligible.

Since, at sufficiently high energy, the Schwarzchild radius is
larger than all microscopic scales beside the radius of the compact
dimensions, we conjecture that the behavior of the elastic
amplitude and at least gross features of multiparticle production
cross sections, can be extracted from the solution of the equations
of classical general relativity.  It is possible that there is
a small
region in impact parameter near to but larger than the Schwarzchild
radius where a more detailed quantum mechanical treatment is necessary.

Thus, in summary we conjecture that most gross features of scattering
at energies much higher than the Planck scale can in fact be
determined by solving classical equations.  This is still a very
involved task.  Even the problem of colliding Aichelburg-Sexl waves
in four flat dimensions is not solved.  For scenarios of
low scale quantum gravity one would have to solve an analogous
problem in a partially compactified space.  Also, one would have to
learn how to extract information about multiparticle amplitudes
from the classical solutions.  Despite the complication, we would
imagine that these problems are amenable at least to numerical
solution.

An important issue which might be clarified by this analysis is
a more precise estimate of the threshold above which our description
of high energy scattering would be expected to hold.  At the
moment we can only say that it should hold sufficiently far above
the Planck scale.  A better estimate of the threshold is crucial
to any attempt to use the properties of black hole production
to constrain theories of low scale gravity.  We would guess that it
is about an order of magnitude higher than the $4+n$ dimensional Planck
mass.

However, even when we reach this threshold, it is not clear that
black hole production will have striking experimental signatures
\footnote{The following paragraphs were a response to questions raised
by E.Witten.}.
The most striking feature of black hole production is of course the
Hawking decay of the final state, which will be nearly at rest in the
center of mass frame.  Unfortunately, for phase space reasons, this
decay will be primarily into KK graviton modes, which are invisible to
all detectors.  Thus, although the final state of black hole decay is
very different from that produced in the perturbative processes
discussed in \cite{kkdetect}, it may not be different in a way that can
be easily measured.  One might hope that at sufficiently high energies,
the Hawking temperature would be so low that KK modes could not be
produced, and we would get a thermal distribution of standard model
particles.   This happens at temperatures where $R_{KK} T < 1$, Since
the Hawking temperature is just the inverse of the Schwarzchild radius,
this is the point at which the Schwarzchild and KK radii cross.  As
noted above, this occurs only at energies larger than the four
dimensional Planck mass.

A more promising signal is the suppression of hard QCD processes.
Complete suppression requires a Schwarzchild radius of order an
inverse GeV.  This occurs at energies of order $(E/M_{4+n} ) \sim
(M_{4+n} / 1 {\rm GeV})^{(n+1)}$.  Even for a six dimensional scenario
with  $M_{4+n} = 1$ TeV, this is $10^{12}$ GeV.  However, suppression of
jets with transverse momenta larger than $M_{4+n}$ will occur as soon
as the threshold for production of black holes is passed.  Furthermore,
the suppression will become more marked with increasing energy, {\it
i.e.} the average transverse momenta of jets should go down with the
energy, precisely the opposite of the QCD expectation.  This question
deserves more detailed study, but we feel confident that a relatively
clean experimental signature will emerge from such a study.
Note that the rate of increase of the Schwarzchild radius with energy
may be measurable in this way, thus providing a direct measurement of
the number of large compact dimensions.

Clearly, all of these studies require the ability to probe a range of
energies up to a few orders of magnitude above $M_{4+n}$, and it is
unclear if anything can be seen in presently planned accelerators.  If
however, evidence for large extra dimensions is found at LHC, then one
would be highly motivated to build a larger machine, which could study
black hole production.

\acknowledgments
We would like to thank G.Horowitz, A.Rajaraman, and E.Witten for useful
discussions.
The work of TB
was supported in part by the DOE under grant
number DE-FG02-96ER40559.
The work of WF was supported in part by the Robert Welch Foundation
and the NSF under grant number
PHY-9219345

\end{document}